\documentclass[manuscript]{acmart}

\usepackage{multirow}
\usepackage{graphicx}
\usepackage{subcaption}
\usepackage{tikz}
\usetikzlibrary{automata,positioning}
\usetikzlibrary{shapes.geometric, arrows}
\usetikzlibrary{calc}
\usetikzlibrary{fit}
\usepackage{pgfplots}
\usepgfplotslibrary{groupplots}
\usepackage{natbib}
\usepackage[most]{tcolorbox}
\usepackage{xcolor} 
\usepackage{mdframed} 
\usepackage{float}
\usepackage{lscape}
\usepackage{longtable}
\pgfplotsset{compat=1.18} 

\AtBeginDocument{%
  }

\setcopyright{acmlicensed}
\copyrightyear{2018}
\acmYear{2018}
\acmDOI{XXXXXXX.XXXXXXX}

\acmConference[Conference acronym 'XX]{Make sure to enter the correct
  conference title from your rights confirmation emai}{June 03--05,
  2018}{Woodstock, NY}
\acmISBN{978-1-4503-XXXX-X/18/06}

\begin{document}

\title{Expert-Generated Privacy Q\&A Dataset for Conversational AI and User Study Insights}

\author{Anna Leschanowsky}
\email{anna.leschanowsky@iis.fraunhofer.de}

\author{Farnaz Salamatjoo}
\email{farnaz.salamatjoo@iis.fraunhofer.de}

\author{Zahra Kolagar}
\authornote{Now with Aleph Alpha}
\email{zarah.kolagar@iis.fraunhofer.de}

\author{Birgit Popp}
\email{birgit.popp@iis.fraunhofer.de}

\affiliation{%
  \institution{Fraunhofer Institute for Integrated Circuits IIS}
  \city{Erlangen}
  \country{Germany}
}

\renewcommand{\shortauthors}{Leschanowsky et al.}

\begin{abstract}
Conversational assistants process personal data and must comply with data protection regulations that require providers to be transparent with users about how their data is handled. Transparency, in a legal sense, demands preciseness, comprehensibility and accessibility, yet existing solutions fail to meet these requirements. To address this, we introduce a new human-expert-generated dataset for Privacy Question-Answering (Q\&A), developed through an iterative process involving legal professionals and conversational designers~\footnote{We make our dataset available through this link \url{https://github.com/audiolabs/Expert-Generated-Privacy-QA-Dataset}}. We evaluate this dataset through linguistic analysis and a user study, comparing it to privacy policy excerpts and state-of-the-art responses from Amazon Alexa. Our findings show that the proposed answers improve usability and clarity compared to existing solutions while achieving legal preciseness, thereby enhancing the accessibility of data processing information for Conversational AI and Natural Language Processing applications.
\end{abstract}


\begin{teaserfigure}
  \includegraphics[width=\textwidth]{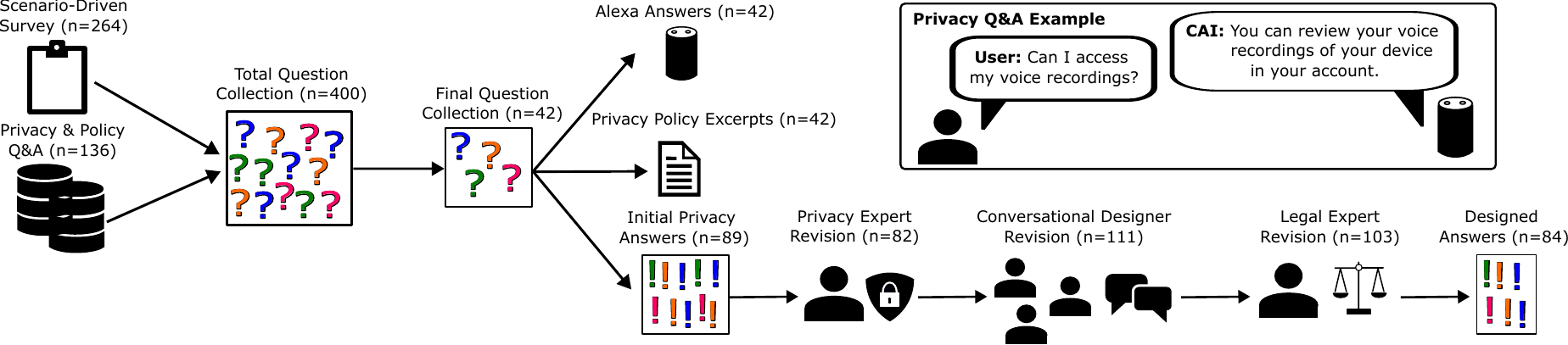}
  \caption{Overview of the data collection process in the study and a privacy question-answering example.}
  \Description{}
  \label{fig:CollectionOverview}
\end{teaserfigure}

\received{20 January 2025}

\maketitle

\section{Introduction}

As Conversational AI (CAI) systems collect, process and store personal data, users need to be informed about how their data is handled. Privacy policies aim to ensure transparency and regulatory compliance (e.g, General Data Protection Regulation (GDPR) in the EU~\cite{european_commission_regulation_2016} or ``Notice and Choice" principle in the US~\cite{ProtectingConsumerPrivacy2012}). However, they are often unread and ineffective for promoting transparency~\cite{mcdonaldCostReadingPrivacy2009, cateLimitsNoticeChoice2010,cranorNecessaryNotSufficient2012,schaubDesignSpaceEffective2018}. Alternatives like privacy labels improve accessibility but lack the detail of natural language policies~\cite{barthUnderstandingOnlinePrivacy2022a, kelley2009nutrition, emami2021informative, reinhardt2021visual}, while machine-readable policies enhance auditability~\cite{morelSoKThreeFacets2020a} but fail to inform users appropriately. Transparency challenges are especially significant for voice-based CAI, as they often lack graphical interfaces. Even simplified textual formats like Amazon Alexa's Privacy FAQ, require users to switch modalities, resulting in limited accessibility~\cite{gautamAlexaWeTrust2022}. Privacy question-answering (Q\&A) could enhance transparency and accessibility in CAI systems by enabling them to respond to privacy-related user questions in natural language (see Figure~\ref{fig:CollectionOverview} for an example).  
Previous research has curated privacy Q\&A corpora based on policy excerpts~\cite{ahmadPolicyQAReadingComprehension2020a, ravichanderQuestionAnsweringPrivacy2019}, but these excerpts remain difficult for non-experts to understand~\cite{ravichanderBreakingWallsText2021}. In addition, questions in these corpora only rarely cover data types relevant to voice-based CAI, such as voice recordings. 

Our research aims to advance privacy Q\&A for CAI by (1) collecting privacy questions for CAI via a scenario-driven survey (Section~\ref{sec:PrivacyQuestions}), (2) analysing the current state-of-the-art by evaluating Amazon Alexa's responses and policy excerpts (Section~\ref{sec:ALexaAnswers} and~\ref{sec:PrivacyPolicyExperts}), (3) developing user-friendly and legally valid answers through collaboration with legal experts and conversational designers (Section~\ref{sec:DesignedAnswers}), and (4) evaluating the three answer types -- Alexa answers, policy excerpts, and our designed answers -- through linguistic analysis and a user study (Section~\ref{sec:LinguisticAnalysis} and~\ref{sec:QualitativeAnalysis}).

\section{Related Work}

Conversational Privacy Bots (PriBots) address the growing need for accessible privacy notices by allowing users to pose privacy-related questions in natural language~\cite{harkousPriBotsConversationalPrivacy, grunewald2023enabling}. 
So far, efforts to support privacy Q\&A have focused on collecting and annotating privacy policy corpora and applying neural networks to extract answers from privacy policies~\cite{wilsonCreationAnalysisWebsite2016, ravichanderQuestionAnsweringPrivacy2019, ahmadPolicyQAReadingComprehension2020a, aroraTaleTwoRegulatory, sathyendraHelpingUsersUnderstanda}. Notable corpora include the OPP-115 Corpus of website privacy policies~\cite{wilsonCreationAnalysisWebsite2016}, the APP-350 Corpus of annotated Android app policies~\cite{zimmeck2019maps} and a bilingual corpus of mobile app policies~\cite{aroraTaleTwoRegulatory}. Other work has focused on collecting privacy Q\&A specific corpora, e.g., the Policy Q\&A corpus~\cite{ahmadPolicyQAReadingComprehension2020a} and the Privacy Q\&A corpus which uses excerpt-based answers identified by legal experts~\cite{ravichanderQuestionAnsweringPrivacy2019}. Despite these efforts, no privacy Q\&A corpus exists for CAI systems. While existing corpora could be adapted, they rarely cover questions and answers specific to CAI systems, e.g., on voice recordings or text transcripts.
Moreover, pre-trained language models used to directly extract answers from privacy policies were found largely unsuitable and face challenges with responding to unanswerable questions~\cite{ravichanderQuestionAnsweringPrivacy2019, ahmadPolicyQAReadingComprehension2020a, ravichanderBreakingWallsText2021}. Thereby, NLP approaches can produce overly technical and lengthy responses that conflict with conversation design principles such as minimization and user-friendly language~\cite{MooreNCF}. At the same time, responses must accurately reflect the essential content of a privacy policy~\cite{harkousPriBotsConversationalPrivacy}. This urges the need for collaboration between conversational designers and legal experts to ensure answers are both comprehensible and legally precise. This is particularly important, given that both comprehensibility and preciseness are aspects of the principle of transparency~\cite{Article29WP2018}, a legal requirement of the GDPR and global data protection regulations~\cite{gunst2021brusselseffect, ProtectingConsumerPrivacy2012}.

\section{Privacy Q\&A Dataset}

Figure~\ref{fig:CollectionOverview} shows our data collection process. We first collect 400 privacy questions and narrow them down to 42 most representative ones (Section~\ref{sec:PrivacyQuestions}). We then construct three types of answers -- Amazon Alexa responses, privacy policy excerpts and human-expert-generated answers (Section~\ref{sec:privacyanswers}).

\subsection{Privacy Questions} 
\label{sec:PrivacyQuestions}

We conducted a scenario-driven survey to collect an initial set of privacy questions (see Appendix~\ref{sec:scenariodriven} for an outline of the survey). Participants were presented with a hypothetical scenario of a privacy breach and asked to generate questions to understand what the CAI system knew about them. Unlike previous privacy Q\&A corpora, which often contain generalized questions about ``data", we aimed to collect questions on nine information types -- commonly collected by smart assistants -- to account for varying degrees of information sensitivity, e.g., voice recordings, contact information and location (see Appendix Table~\ref{tab:datatype} for details)~\cite{smartassistantdatacollection, schomakersInternetUsersPerceptions2019}. In total, we collected 264 questions from 11 internally recruited participants and categorized them into six data practices categories (see Appendix Table~\ref{tab:datapractice} for details) using existing privacy Q\&A annotation schemes~\cite{wilsonCreationAnalysisWebsite2016, aroraTaleTwoRegulatory}. People were not compensated for participation and we did not collect any demographic data. To account for the small sample size and broaden the scope of questions, two of the authors added questions from Privacy Q\&A and Policy Q\&A corpora, resulting in a dataset of 400 questions covering nine information types and six data practice categories~\cite{ahmadPolicyQAReadingComprehension2020a, ravichanderQuestionAnsweringPrivacy2019}.

To generate answers to these questions that are both legally precise and comprehensible, we pursued
an iterative experts-in-the-loop process and recruited legal and dialogue design experts~\cite {kolagar2023experts}. As this expert-in-the-loop approach is time-intensive and experts are difficult to recruit, we selected the most representative question per information type and data practice using Semantic Textual Similarity (STS) with Sentence-BERT~\cite{reimers19sentence}. This minimized the workload for experts, while at the same time maximizing the diversity of the resulting Q\&A dataset. We further excluded questions that could not be sufficiently answered, given the available privacy policy, resulting in 42 final questions relatively equally distributed across data practices and with a stronger focus on voice recordings (see Appendix Table~\ref{tab:datapractice} and~\ref{tab:datatype}).

\subsection{Privacy Answers} 
\label{sec:privacyanswers}

Unlike previous research that focused on extracting policy excerpts, we aim to create human-expert-generated user-friendly and legally precise answers to privacy questions. For evaluation, we collect three different types of answers: (1) Amazon Alexa responses (Section~\ref{sec:ALexaAnswers}), (2) excerpts from Amazon's privacy policy and related sources (Section~\ref{sec:PrivacyPolicyExperts}), and (3)  answers created by conversational designers and legal experts (Section~\ref{sec:DesignedAnswers}). 

\subsubsection{Privacy Answers by Amazon Alexa} 
\label{sec:ALexaAnswers}

To evaluate the privacy question-answering capabilities of CAI, we queried Amazon Alexa using the Amazon Alexa App\footnote{Alexa Answers in the dataset were collected in January 2023. To ensure the validity of the answers, we re-evaluated answers in January 2025 and observed no drastic changes in responses.}. We prompted the system to repeat what it heard after posing each question to ensure correct speech recognition and asked questions multiple times to account for variabilities in response behaviour. Despite multiple trials with different people and voices, no response was obtained for five out of 42 questions. We denote these cases with ``No answer" in the dataset (see Appendix~\ref{sec:CategorizationAlexa} for a detailed overview). Most privacy questions could not be answered and relied instead on fallbacks like general excuses or redirection to the help section. Only four questions were ``correctly'' answered by providing control options or privacy information. Notably, questions on voice recordings could be answered, possibly due to public outcry and media coverage on the topic over the last years~\cite{alexaoutcry}. 

\subsubsection{Privacy Policy Excerpts} 
\label{sec:PrivacyPolicyExperts}

In line with previous studies that had legal experts extract answers from privacy policies~\cite{ravichanderQuestionAnsweringPrivacy2019}, we extracted excerpts from the latest Amazon Europe privacy policy~\cite{AmazonAlexaPrivacy} and additional resources like the Alexa and Alexa Device FAQs\footnote{\url{https://www.amazon.com/-/de/gp/help/customer/display.html?nodeId=201602230}} and Alexa Features Help\footnote{\url{https://www.amazon.com/-/de/gp/help/customer/display.html?ref_=hp_bc_nav&nodeId=G201952240}}. These FAQs and help pages were particularly helpful for answering user choice and control related questions. While we explored advanced tools for privacy policy excerpt extraction, we decided on a manual approach. First, tools for extraction like ``PriBot''~\cite{harkousPriBotsConversationalPrivacy} or the machine-annotated privacy policies provided by the Usable Privacy Policy Project\footnote{\url{https://explore.usableprivacy.org/}} were either unavailable to the authors at the time of the study or used outdated privacy policies exclusively applicable to the US. Given our focus on the European privacy policy following the enforcement of GDPR, these tools did not meet our needs. Second, the use of Large Language Models (LLMs), e.g., OpenAI's GPT-4 model, with template prompting, was ineffective as the model produced untruthful and unusable excerpts, relying more on general knowledge than the provided documents.

\subsubsection{Privacy Answers Designed by Experts}
\label{sec:DesignedAnswers}

To create comprehensible and legally precise answers, we used an iterative process with legal experts and conversational designers as shown in Figure~\ref{fig:CollectionOverview}~\cite{kolagar2023experts}. Drawing from the previously extracted privacy excerpts (see Section~\ref{sec:PrivacyPolicyExperts}), two researchers independently curated an initial set of privacy answers with two or more answers per question. We explicitly derived the answers from the extracted excerpts to ensure coherence and consistency across answer types. First, we consulted with a trained privacy technologist (CIPT~\footnote{IAPP Certified Information Privacy Technologist (\url{https://iapp.org/certify/cipt/})}) to ensure that responses were privacy-preserving and omitted those that did not satisfy the requirements. Following this, three conversational designers revised the answers focusing on their comprehensibility, conversational flow, and user-friendliness. The experts had diverse backgrounds in conversation design, voice user interface design and linguistics with professional experience ranging from one to five years. Finally, to ensure legal validity, we revised answers based on a legal expert's feedback specializing in data protection and IT law with more than 5 years of experience. Each question was answered in more than one way resulting in one-to-many mapping for each question. This makes sense, as privacy questions can be answered validly by multiple responses, maintaining both legal preciseness and comprehensibility. Given the final set of 103 designed answers -- a set too large to be evaluated in the user study -- we again leveraged STS~\cite{reimers19sentence} to choose the two most distinct answers per question, i.e., ``Designed Answer 1'' and ``Designed Answer 2''.


\section{Linguistic Analysis of Privacy Answers}
\label{sec:LinguisticAnalysis}

To understand how answers differed objectively, we conducted a linguistic analysis focusing on readability and lexical diversity, characteristics that have been used before to evaluate privacy policies (see Table~\ref{tab:linguisticanswers} for details)~\cite{milne2006longitudinal, cadogan2004imbalance, hamid2023privacylens}. 

\begin{table}[!ht]
    \centering
    \begin{tabular}{lcccc}
        \toprule
        Linguistic Metrics & Alexa Answers & Excerpts & Designed Answers 1 & Designed Answers 2 \\
        \midrule
        \multicolumn{5}{l}{Readability}  \\
        \midrule
        Flesch-Kincaid Grade Level & 3.5 & 11.3 & 8.4 & 8.0 \\
        Flesch Reading Ease & 87.2 & 47.6 & 62.6 & 66.32 \\
        Linsear Write Index & 5.0 & 12.75 & 10.4 & 8.4 \\
        \midrule 
        \multicolumn{5}{l}{Lexical Diversity} \\
        \midrule 
        Unique Word Count & 10.0 & 46.5 & 22.0 & 23.5 \\
        Type-Token-Ratio & 1.0 & 0.67 & 0.82 & 0.81 \\
        Measure of Textual Lexical Diversity & 16.0 & 56.75 & 45.17 & 37.99 \\
        \bottomrule
    \end{tabular}
    \caption{We show the median results for the linguistic metrics for each answer type.}
    \label{tab:linguisticanswers}
\end{table}

\noindent\textbf{Readability} Our readability analysis using the Flesch-Kincaid grade level and Flesch reading ease~\cite{flesch1949art} shows that Alexa answers score lowest in readability while excerpts score highest, requiring a grade of 11 on the Flesch-Kincaid scale. However, the low score for Alexa answers is primarily due to brief responses, e.g., ``Sorry, I don't know that.''. Although small differences exist between both sets of designed answers, they are more readable than the excerpts, making them easier for people to understand. Further, we calculated the Linsear Write Readability Index, originally developed to improve government writing~\cite{o1966gobbledygook}. Importantly, the index calculation differs from the Flesch-Kincaid measures as it considers both sentence length and the number of words with three or more syllables. Again, Alexa answers are the easiest while the excerpts contain more difficult words. However, the index also shows a difference between the two designed answer categories, indicating that the first set uses longer words compared to the second set. Despite the iterative process, expert-created answers show some degree of variability if assessed objectively.

\noindent\textbf{Lexical Diversity} As a proxy for technical and legal jargon which contributes to unique and low-frequency words, we compute lexical diversity as a measure of how many different words are used in a text. While low-frequency words can be crucial to understanding privacy policy texts, they can make comprehension challenging~\cite{hiebertunique}. While we compute several metrics such as Type-Token-Ratio (TTR) and measure of textual lexical diversity (MTLD), we rely on MTLD as our main measure as it was found to be least affected by text length~\cite{zenker2021investigating, koizumi2012effects}. In line with readability scores, we find that Alexa Answers are the least diverse while excerpts are the most. Our designed answers show results in between with the first set showing higher MTLD scores in line with the Linsear Write Readability Index results. 

\section{User Study}
\label{sec:QualitativeAnalysis}

\subsection{Experimental Design}

For subjective human evaluation of generated responses, we adopted a mixed-method approach to capture quantitative and qualitative insights on four evaluation criteria, i.e., answer quality, usability, difficulty and lawyerliness, previously used to assess legal documents~\cite{Martinez2023}. Participants rated answers on how well they met their information needs \textbf{(Answer Quality)}, how much they sounded appropriate in style and tone for legal information \textbf{(Lawyerliness)}, how useful they were given the question \textbf{(Usability)} and how complex and difficult they perceived them \textbf{(Complexity/Difficulty)}.

We conducted semi-structured online interviews and recorded and transcribed videos. Best-Worst Scaling (BWS) was chosen to gather quantitative insights~\cite{Cohen2003MaximumDS, louviere2015best}. In each round, participants were presented with five questions, including information-seeking questions, yes/no questions and statements, and covering different information types, i.e., name, contacts, time zone, calendar, and voice recordings (see Appendix Table~\ref{tab:QuestionAnswerUserStudy} for details on questions and answers). 

Participants completed the study in four rounds, each measuring one specific metric. The study was displayed using Amazon Mechanical Turk Sandbox (see Appendix~\ref{fig:AMTInterface} for the interface). In each round, participants rated five questions, each paired with three of the four answer types -- Alexa answers, policy excerpts, and designed answers 1 and 2. In total, participants rated 20 question-answer trials. We randomized the order of questions, answers, and metrics to avoid choice bias. Participants first performed best-worst scaling and rated answers based on the evaluation criteria, then, they provided explanations for their choices. Finally, they were invited to offer general feedback.

\subsection{Participants and Ethical Considerations}

Participants, two males and two females with an average age of 30, had expertise in linguistics which provided valuable insights into how specific language features can influence comprehension and user experience. People were internally recruited and not compensated for participation. Before the interview, we obtained verbal consent for video recording the interview sessions and ensured that their data remained confidential. 

\subsection{Quantitative Results} 
\begin{figure}[!hbt]
    \centering
    \resizebox{0.8\textwidth}{!}{\input{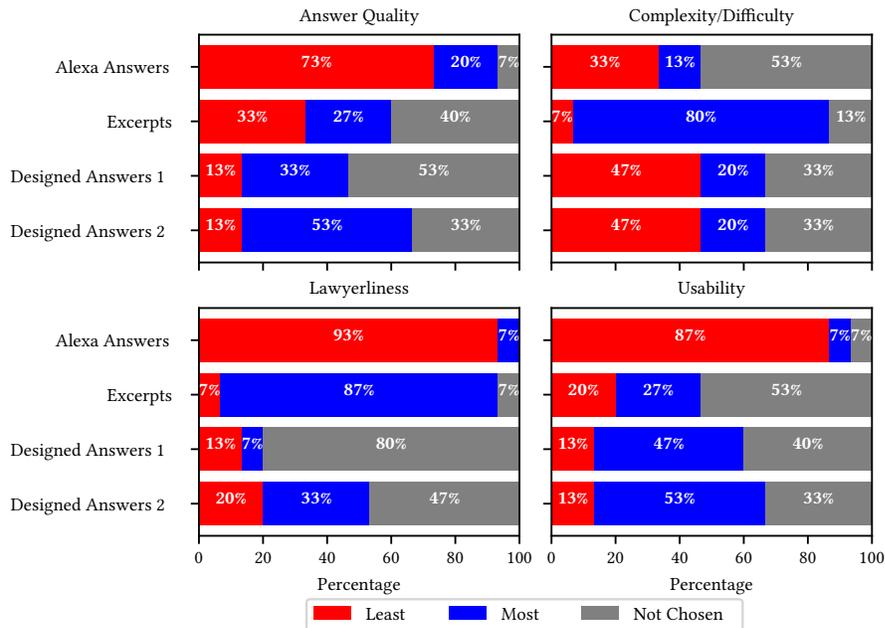}} 
    \caption{Quantitative results comparing the four answer types across the metrics of quality, difficulty, usability, and lawyerliness.}
    \label{fig:percentages}
\end{figure}

Figure~\ref{fig:percentages} shows the percentage of trials where each answer type was rated the best, worst or was not chosen per evaluation criterion. Excerpts were rated as the most difficult to understand, while designed answers 1 and 2 were considered the least difficult in 47\% of cases they were shown to participants. Excerpts were also judged the most lawyerly in 93\% of cases, whereas Alexa answers were perceived as the least lawyerly. For quality and usability metrics, designed answer 2 received the highest ratings, being marked as the best in 53\% of trials for both metrics. In contrast, Alexa answers were rated the lowest, with 73\% of trials identifying them as having the least quality and 87\% as having the least usability.

\subsection{Qualitative Analysis}

\subsubsection{Coding Strategy}
We explored participants' perceptions of privacy answers by identifying the recurring themes through inductive coding~\cite{thomas2006inductive}. This process involved creating categories of codes based on user feedback and helped capture both positive and negative feedback for each answer type (see Appendix Table~\ref{tab:1} for the codebook). The extracted codes reflect stylistic aspects including informal style, formal style, and instructional clarity, as well as linguistic aspects like semantic complexity, rich syntactic structure, content comprehensiveness, and keyword significance. 

In addition, the codes offered insights into how participants perceived each metric. Figure~\ref{fig:Frequency of Codes} in the Appendix displays the top three most frequently used codes for each metric. For instance, participants associated low-quality and low-usable responses with irrelevant answers, while high-quality and usable responses were linked with straightforward and comprehensive responses. 

\subsubsection{Qualitative Results} \hfill

\noindent\textbf{Alexa Answers} were frequently perceived as irrelevant and lacking sufficient details. Specifically, P1 stated that \textit{``there is nothing on the calendar for the next 30 days is unrelated [to the question]''} and P3 noted that \textit{``it doesn't provide any information at all''}. However, the simplicity of language, short sentences and simple syntactic structures in these answers were positively received by P1 and P2. 


\noindent\textbf{Privacy Policy Excerpts} were praised for being comprehensive and giving specific responses. Participants stated that their sentence length contributed to the higher quality and more lawyerly tone. As P3 remarked, \textit{``The reason why this one is the most lawyerly is that it's quite verbose because they are being very explicit and making sure that every detail is written down''.} On the other hand, lengthy instructions and difficult semantics and terminology posed challenges for P1, P3, and P4, because they contributed to the complexity of responses. For instance, P3 highlighted \textit{``It provides like good information but it's too long, too complicated''.}   





\noindent\textbf{Designed Answer 1} were liked for being straightforward, making the answers more useful. One factor which contributed to their ease of understanding was the use of imperative verbs rather than modal verbs. P1 mentioned their preference, pointing out \textit{``I chose this one because it says do this instead of `you can' with the auxiliary.''}. Imperative constructions decreased sentence complexity and offered direct instructions to the user. In contrast, from a negative perspective, these answers were confusing because they lacked specificity, which diminished their clarity and legal tone. All Participants stated that insufficient information would lead to lower quality and usability of answers. P4 commented \textit{``It's not informing me about anything and is very evasive and vague''.} 




\noindent\textbf{Designed Answer 2} mostly received positive feedback as they were perceived as straightforward and succinct, giving sufficient details without leading to complexity. According to the participants, this feature added to the usability and lawyerliness of the answers, while also reducing the level of difficulty in comprehension. In particular, keywords in some of the answers were considered helpful by all participants in directing their decision-making process. For example, the answer ``You can give permission to Alexa to add, delete, or update your Calendar events'' highlighted the user's authority to grant or withhold consent. P3 appreciated this characteristic, saying, \textit{``It offers a next step of how you might be able to do it.''} Similarly, P1 remarked \textit{``Because of the word permission, I feel like it gives more insight into that. I gave consent to this in a sense.''}. This highlights the semantic feature of the sentence with the word ``permission'' shaping the user's interpretation. Participants criticized a few of these answers for having long sentences, making them harder to retain and follow. The combination of sentence length and the placement of the answer at the end of the sentence hindered comprehension. P2 found it frustrating, that \textit{``You can obtain the time zone information with my usage of Amazon service, but therefore the answer comes at the end.''} Likewise, P1 complained about the answer length, noting \textit{``It's very lengthy, and I don't want my interaction. If I want Alexa to answer the question of can I do this, I prefer not to receive a very lengthy answer.''} Moreover, technical terms such as ``cloud'' were found not to be intuitive for all users and can potentially add to the complexity of the sentence.


%


\section{Discussion and Future Work}

Our linguistic analysis and subjective evaluation show that answers designed by experts differ from current state-of-the-art (Alexa responses and Privacy Policies) and are perceived more positively. The codes provide a framework for mapping the stylistic and linguistic features of the responses to the subjective understanding of the answers. Designed answers are perceived as higher in overall quality and more usable than alternatives. Privacy excerpts are rated as the most lawyerly, which may reflect the difficulty of understanding them. Designed answers are perceived as more lawyerly than responses given by Alexa, yet less complex and easier to understand. In addition, our analysis provides insights into metric definition and understanding (Figure~\ref{fig:Frequency of Codes}). Answer quality and usability were evaluated similarly and associated with the same stylistic and linguistic aspects, suggesting that future research could focus on only one of the two aspects. Our inductively extracted codes show that participants identified specific characteristics when describing high-quality and usable answers, as well as low-quality and unusable ones. Future work could align these characteristics with objective metrics used to evaluate state-of-the-art NLP applications~\cite{srivastava2024towards}. This could help scale our iterative approach in generating comprehensible, legally precise and accessible answers. 

We believe this dataset and the approach of creating gold-standard answers to privacy questions in an iterative expert-in-the-loop approach is a significant step forward compared to the current state-of-the-art Privacy Q\&A datasets. There are three main reasons for this which align with the legal principle of transparency~\cite{Article29WP2018}, underlying international data protection legislations~\cite{coe108plus2018,ccpa2018,lgpd2018}: (1) Accessibility: ``The `easily accessible' element means that the data subject should not have to seek out the information''~\cite{Article29WP2018}. For conversational technology, the most accessible way to provide information is ``as an answer to a natural language question''~\cite{Article29WP2018}. Note, that extractions of privacy policy excerpts can not generally be considered a conversational response as they often do not represent how people speak and respond in conversations. (2) Comprehensibility: ``clear and plain language must be used''~\cite{Article29WP2018}. This is not typically the case for legal texts -- \citet{Martinez2023} find that even legal professionals judge legal texts to be poorly comprehensible. The current state-of-the-art of extracting legal language from privacy notices perpetuates incomprehensibility of data processing information. (3) Preciseness: ``the data subject [...] should not be taken by surprise [...] about the ways in which their personal data has been used''~\cite{Article29WP2018}. Privacy notices are often written in vague, complex and confusing language to obscure the true extent of data processing and give companies broad discretion in handling personal data~\cite{solove2013privacy}. For the creation of our dataset, we recruited multiple, independent (not associated with Amazon) and multi-disciplinary experts into an iterative exchange with the goal to assure both legal precision as well as comprehensibility. Thus we depart from the -- possibly obfuscated -- privacy policies, enrich and interpret the original privacy policy with the help of independent experts to provide answers as truthfully as possible based on the available information.

\section{Limitations}

We acknowledge several limitations in our study. First, the small participant sample in the question collection -- despite supplementing it with questions from existing privacy Q\&A corpora -- and the user study limits the generalizability of our findings. A qualitative study with a larger and more diverse sample, including naive users, could provide deeper insights into answer quality. Second, comprehensibility needs can vary across user groups. For instance, children may require simpler language while users with higher privacy literacy may benefit more from extended privacy Q\&A systems~\cite{ravichanderBreakingWallsText2021}. Future research could instruct conversational designers to construct answers for specific user groups. Finally, legal experts can interpret policy language differently~\cite{reidenberg2015disagreeable, ravichanderBreakingWallsText2021} and thus, may differ in their views on what constitutes legally precise formulations. Our designed answers are likely to reflect such variations, and different experts might have created alternative privacy answers. 

\section{Conclusion}

Our study presents a first step to more comprehensible, accessible, and legally precise privacy answers in CAI systems. Through our expert-driven process, we designed privacy answers that strike a balance between user-friendliness and legal accuracy, addressing gaps in the comprehensibility of existing privacy notices. Quantitative and qualitative evaluations show that our designed answers improve usability and clarity compared to existing solutions while maintaining a lawyerly tone where appropriate. Our approach aligns with key principles of transparency in the GDPR, and international transparency principles, ensuring that users are not left confused or misinformed about data practices. Future research should explore scaling this iterative expert process, accounting for different user needs and contexts, while continuing to focus on accessibility and legal compliance in privacy Q\&A datasets for CAI systems.

\begin{acks}
We would like to thank the conversational designers and legal experts for their time and valuable contribution to the dataset. 
\end{acks}

\bibliographystyle{ACM-Reference-Format}
\bibliography{sample-base}

\appendix

\newpage 
\section{Scenario-Driven Survey for Privacy Question Collection}
\label{sec:scenariodriven}

\begin{figure*}[!h]
\begin{mdframed}[
    linecolor=black,
    linewidth=1pt,
    roundcorner=10pt,
    backgroundcolor=white,
    innerleftmargin=15pt,
    innerrightmargin=15pt,
    innertopmargin=15pt,
    innerbottommargin=15pt,
]
\small
\textbf{My Perfect Voice Assistant} \\
You and your husband have recently bought a voice assistant for your house called ‘Silvia Home’. 
Since you have started to use it on a daily basis, you have realized how efficient and advanced the system is, and how all these mundane tasks you had to do before, are now a simple command away!

\vspace{5pt} 

\textbf{My Mom and Silvia} \\
Yesterday, your mom was giving you a visit and after you told her about ‘Silvia’, she became curious on how the system performs, so she started to use the device in the kitchen to set the timer for when she needs to take the food out of the oven, and since she was home alone, she decided to talk to Silvia about her life...

\vspace{5pt} 

A few days later ...

\vspace{5pt}

\textbf{How much does Silvia know about us?} \\
As mother’s day was approaching, you thought you could order some flowers and have them delivered to your mom’s home. When you asked Silvia to order roses for mother’s day, it said, ``but your mom doesn’t like roses.'' You tried to ignore that comment but got concerned when Silvia said, ``but your mom lives on Westfall Avenue.'' These comments made you uncomfortable and curious. You wondered ``how much does Silvia know about me?''

\vspace{5pt} 

\textbf{Task Description} \\
We would like you to imagine that you are Maria who is by now concerned about how much `Silvia', the voice assistant, knows about her and her life. Below, we have provided a set of possible topics you can ask Silvia about.  
Your task is to formulate questions for Silvia to learn how much she knows about you. Please keep each topic in mind while formulating your questions to Silvia. 

\end{mdframed}
\end{figure*}

\section{Data Practice and Information Type Categories Covered by the Collected Questions}

\begin{table}[H]
    \begin{tabular}{p{6cm}|p{3.75cm}|p{3.75cm}}
        \toprule
        Data Practice & Proportion of Total Question Collection (N=400) & Proportion of Final Collection (N=42)\\ 
        \midrule
        First-Party Collection/Use - Information Type & 50\% & 17\%\\
        User Rights and Choice/Control & 18\% & 19\%\\
        Third-Party Collection/Use &  13\% & 21\%\\
        First-Party Collection/Use - Purpose & 8\% & 24\%\\
        Data Security and Retention & 9\% & 14\%\\
        Privacy Policy & 2\% & 5\%\\
        \bottomrule
        \end{tabular}
        \caption{Data practice categories and their proportion in the total question collection and the final question collection for the expert-generated Privacy Q\&A dataset. We emphasize the use of a more balanced question collection for the dataset creation.}
    \label{tab:datapractice}
\end{table}

\begin{table}[H]
      \centering
        \begin{tabular}{l|l|l}
        \toprule
        Information Type & Proportion of Total Question Collection (N=400) & Proportion of Final Collection (N=42)\\ 
        \midrule
        Location & 28\% & 12\%\\
        Contacts & 12\% & 14\%\\
        Age & 12\% & 5\%\\
        Voice Recordings & 11\% & 17\%\\
        Name &  11\% & 14\%\\
        Calendar & 10\% & 12\%\\
        Email & 6\% & 7\% \\
        Phone Number & 5\% & 9.5\%\\
        Time Zone & 5\% & 9.5\%\\
        \bottomrule
        \end{tabular}
    \caption{Information type categories and their proportion in the total question collection and the final question collection with a stronger focus on questions related to the topic of voice recordings.} 
    \label{tab:datatype}
\end{table}

\section{Categorization of Alexa Answers}
\label{sec:CategorizationAlexa}

\begin{table}[H]
    \centering
    \begin{tabular}{l|c|p{9cm}}
        \toprule
        Response Type & Proportion (n=42) & Example \\
        \midrule
        Excuse & 28.6\% & Sorry, I don't know that.\\         
        ``No answer" & 11.9\% & - \\         
        Online Reference & 11.9\% & From Redflash.com: Amazon does not provide your email or phone numbers to third-party sellers. \\ 
        Forwarding & 9.5\% & For help with that question go to the help and feedback section of the Alexa App. \\
        Account Information & 9.5\% & I am not sure who's speaking but you're in X account. To teach me to recognize your voice just say ``Learn my voice" \\
        Calendar Information & 9.5\% & There's nothing on the calendar for the next 30 days \\
        General Information & 9.5\% & This device's time zone is Central European Time. \\
        Control \& Choice & 7.2\% & Okay, I'll delete any recordings from the last ten minutes. \\
        Privacy Information & 2.4\% & Your data allows me to respond to you and helps me learn from our interactions so that I can better help you. For example, the more variation in spoken language I learn, the better I'll be at understanding what you mean. You can find settings for how your data is used by visiting amazon.co.uk/alexaprivacysettings or the Privacy Section of the Alexa app. \\  
    \bottomrule
    \end{tabular}
    \caption{Categorization of answers provided by Amazon Alexa to the questions in the final collection and their proportion. We find that the majority of answers provide an excuse. Information on control and choice or privacy information was specifically provided for questions related to the voice recordings.}
    \label{tab:AlexaAnswers}
\end{table}

\section{Questions and Answers in the User Study}

\begin{landscape}
\small
\centering
    \begin{longtable}{|p{0.15\textheight}|p{0.2\textheight}|p{0.3\textheight}|p{0.25\textheight}|p{0.25\textheight}|}
        \toprule
        Question & Alexa Answer & Privacy Policy Excerpt & Designed Answer 1 & Designed Answer 2 \\
        \toprule
        I want to change my name. & You can find and update my wake word in the Alexa app. Select the devices icon, select your device and then select Wake Word. & You can add or update certain information on pages such as those referenced in the Information You Can Access section. When you update information, we usually keep a copy of the previous version for our records. & Visit the Alexa Profile Settings on your Alexa App to edit information on your name. & You can manage your personal information under your profile in your Amazon settings.\\
        Do they keep my contacts' information? & Sorry, I don't have an answer for that. & When you register for Alexa Communication on your Alexa app, you will be asked to import your contacts from your device, which will then appear as contacts in your Alexa app. When you open the Alexa app, your contacts are auto-updated from your tablet or mobile phone. You may disable contact import in the Alexa app. If contact import is disabled, contacts you previously imported from your device will be deleted. You may also manually add contacts to your Alexa contact list in the Alexa app or at https://alexa.amazon.com. You can also delete each contact you manually added by selecting the applicable contact card. & When you register for Alexa Communication on your Alexa app, you will be asked to import your contacts from your device, which will then appear as contacts in your Alexa app. & If you use Alexa Communication to communicate with friends and family via Alexa, you will be asked to import your contacts from your device. They will be updated when you open the Alexa App. To delete contacts you can disable contact import or delete them manually by selecting the corresponding contact card in your Alexa App. \\
        Can the assistant pull information from my phone's calendar? & There's nothing on the calendar for the next 30 days. &To have Alexa add, delete, or update your calendar events, link your calendar in the Alexa app. Note: When you link your calendar account, a copy of your calendar is stored in the cloud so you can access it using Alexa. Content from your calendar might be available to anyone using your Alexa devices. &You can link your calendar account to your Alexa in the Alexa App. This will allow anyone with access to your Alexa devices to manage events. The Alexa Calendar is stored in the cloud. &You can give permission to Alexa to add, delete, or update your Calendar events and also link your calendar in the Alexa app. If anyone else is using your Alexa devices, the content of your calendar might be available to them. \\
        Do you need to know my time zone?  & This devices time zone is Central European Time. & We automatically collect and store certain types of information about your use of Amazon Services, including information about your interaction with products, content, and services available through Amazon Services. Examples of the information we collect and analyze include: version and time zone settings. &We collect your Time Zone automatically to provide you with a more effective experience. & When you interact with our products or services we might automatically receive information on your time zone. In addition, your device's location is used for providing time or weather information. You can change your location at any time by editing your device location on your Alexa App.\\ 
        Who can access my voice recordings? & For help with this question go to the help and feedback section of the Alexa app.& Using your browser, you can listen to Alexa recordings in the Review Voice History section of Alexa Privacy Settings (amazon.com/alexaprivacysettings). Using the Alexa app, you can listen to Alexa recordings by going to the menu and selecting Settings > Alexa Privacy > Review Voice History. & You can access your voice recordings in your Voice History in the Alexa Privacy Settings. & You can review voice recordings of your device in your account. \\
    \bottomrule
    \caption{Questions and Answers displayed in randomized order in the user study.}
    \end{longtable}
    \label{tab:QuestionAnswerUserStudy}
\end{landscape}

\section{Codebook Derived from the Interviews}

\begin{table}[!htb]
    \centering
    \small
    \begin{tabular}{p{3cm}p{4cm}p{7cm}}
       \toprule
        Codes & Definitions & Examples \\
        \midrule
        \multicolumn{3}{l}{Stylistic}  \\
         \midrule
        Informal Style & conversational or casual speech & This devices time zone is Central European Time\\
        Formal Style & more professional language that avoids directly addressing the user & You can add or update certain information on pages such as those referenced in the Information You Can Access section. When you update information, we usually keep a copy of the previous version for our records\\
        Lengthy Instruction & answers with excessively detailed steps & Using your browser, you can listen to Alexa recordings in the Review Voice History section of Alexa Privacy Settings (amazon.com/alexaprivacysettings). Using the Alexa app, you can listen to Alexa recordings by going to the menu and selecting Settings > Alexa Privacy > Review Voice History\\
        Straightforward & quite short sentences providing quick and direct answers & Visit the Alexa Profile Settings on your Alexa App to edit information on your name\\
         \midrule 
        \multicolumn{3}{l}{Linguistic} \\
        \midrule 
        Difficult Semantics & technical and sophisticated terminology & To have Alexa add, delete, or update your calendar events, link your calendar in the Alexa app. Note: When you link your calendar account, a copy of your calendar is stored in the cloud so you can access it using Alexa. Content from your calendar might be available to anyone using your Alexa devices\\
        Irrelevant & the sentence does not answer the question at all & User: Can the assistant pull information from my phone's calendar? Alexa: There's nothing on the calendar for the next 30 days \\
        Simple Semantics & responses free of technical or convoluted language & You can give permission to Alexa to add, delete, or update your Calendar events and also link your calendar in the Alexa app. If anyone else is using your Alexa devices, the content of your calendar might be available to them\\
        Rich Syntactic Structure & complex and varied grammatical constructions, aimed at providing detailed information & We automatically collect and store certain types of information about your use of Amazon Services, including information about your interaction with products, content, and services available through Amazon Services. Examples of the information we collect and analyze include: version and time zone settings\\
        Comprehensiveness & the answers which in terms of content were long, thorough, and detailed & To have Alexa add, delete, or update your calendar events, link your calendar in the Alexa app. Note: When you link your calendar account, a copy of your calendar is stored in the cloud so you can access it using Alexa. Content from your calendar might be available to anyone using your Alexa devices\\
        Keyword Significance & specific words within sentences shape participants' perceptions and influence their decision-making & You can review voice recordings of your device in your account\\
        \bottomrule
    \end{tabular}
    \caption{Codebook presenting the defined codes, their definitions derived from interview feedback, and examples taken from the dataset.}
    \label{tab:1}
\end{table}

\section{Interface of Amazon Mechanical Turk Sandbox Displaying Interview Questions}

\begin{figure}[H]
    \centering\includegraphics[width=0.75\linewidth]{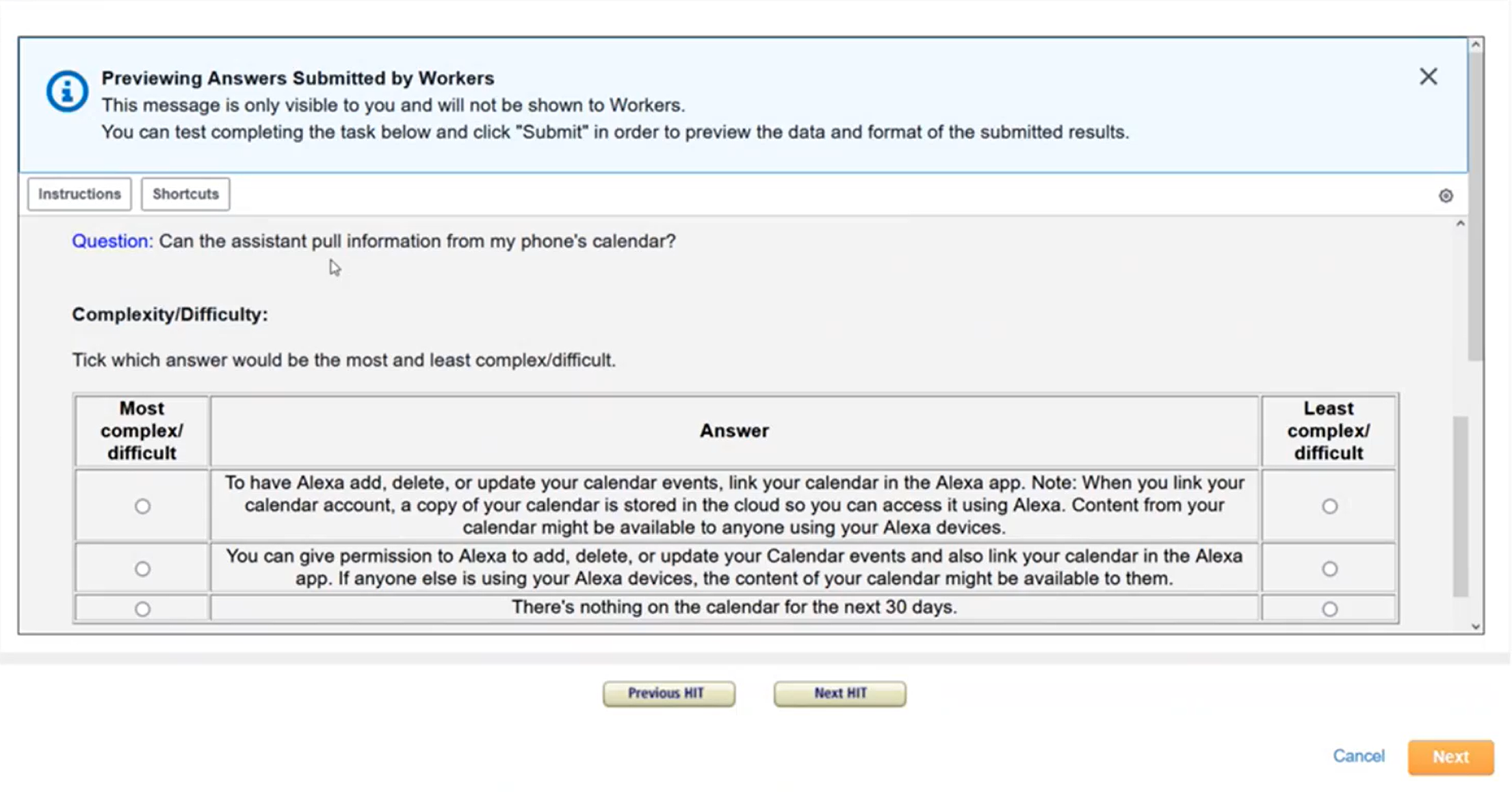}
    \caption{Example of the questions interface in the Amazon Mechanical Turk Sandbox as seen by participants during the interviews.}
    \label{fig:AMTInterface}
\end{figure}

\section{Frequency of Top Three Codes by Metric}

\begin{figure}[H]
    \centering
    \includegraphics[width=0.75\linewidth]{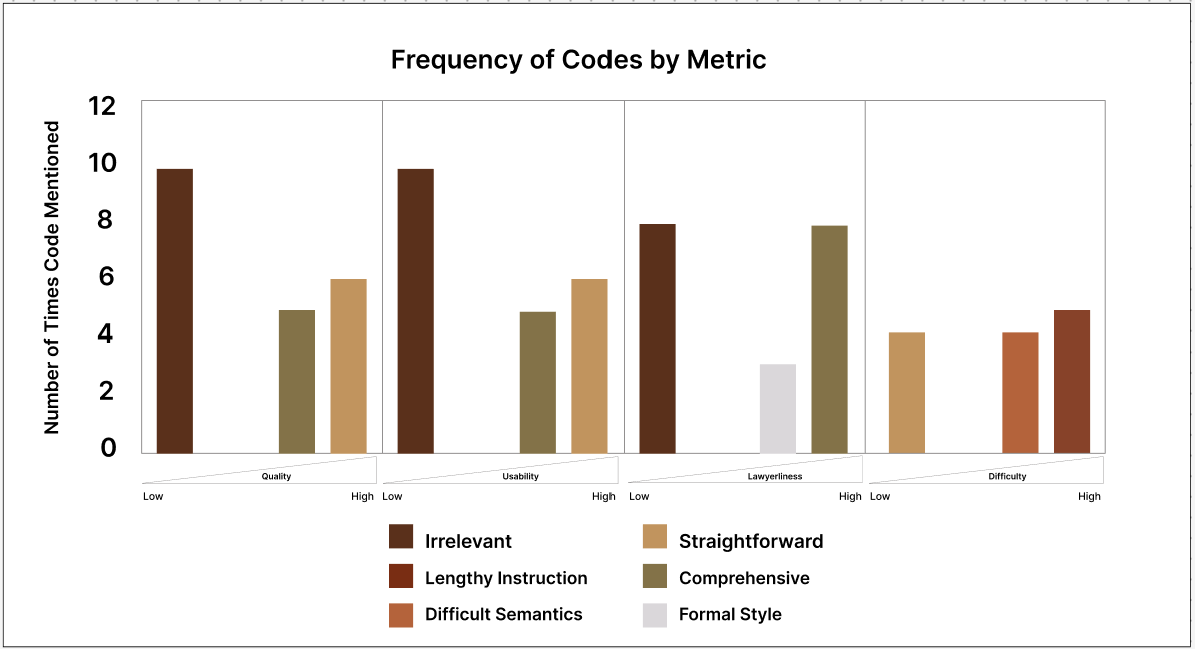}
    \caption{The figure displays the three most frequently occurring codes in the interview feedback for each metric, demonstrating which aspects were most noticeable and important to participants during their subjective evaluation of the answers.}
    \label{fig:Frequency of Codes}
\end{figure}

\end{document}